# Dodecagonal spherical quasicrystals

I.A. Shevchenko, O.V. Konevtsova and S.B. Rochal

*Faculty of Physics, Southern Federal University, 5 Zorge str., 344090 Rostov-on-Don, Russia*

We argue that 2D dodecagonal spherical quasicrystalls (QCs) will be discovered in the nearest future and investigate how the planar QC order becomes compatible with the spherical geometry. We show that the appearance of curvature-induced topological defects and low-energy structural rearrangements are sufficient to obtain the regular spherical QC structures. Minimization of total energy required for the order reconstruction determines the number of topological defects, which are located near the vertices of snub cub or icosahedron in dependence on type of the initial dodecagonal order.

PACS numbers: 68.65.-k, 61.44.Br, 62.23.St, 02.40.Re

A large number of 2D nano- and micro - structures with a spherical topology possess a local hexagonal order which is compatible with the global icosahedral or tubular geometries. There are three widespread types of the hexagonal orders with one, two and six particles per unit cell, respectively. The simplest one is locally implemented in spherical colloidal crystals and colloidosomes [1-4]. The next arrangement is demonstrated by the graphene structure being a useful parent phase to explain the structural and physical properties of more complicated systems with the same local organization typical of fullerenes and carbon nanotubes [5-6]. The local hexagonal order with six proteins in the cell underlies Caspar and Klug theory [7-8] which is the basis of almost all the works devoted to the physics of spherical viruses.

In contrast, 2D spherical QCs are very rare in the nature. As far as we know only some spherical viral capsids possess the quasicrystalline order. The first tiling models of polyoma viral capsid was proposed in Ref. [9]. In addition, several capsids were recently interpreted as 2D spherical structures with a local chiral pentagonal quasicrystalline organization of proteins [10,11]. However, the discovery of the quasicrystalline order in other spherical systems is very likely. In the past decade many dodecagonal quasicrystalline systems were found. These are dendritic micelles [12], star and tetrablock terpolymer melts [13,14], diblock copolymer [15], and surfactant micelles [16]. The discovered QCs are soft like well-known spherical colloidal crystals. Therefore we believe that soft dodecagonal spherical QCs can be obtained in the nearest future. We are not alone in this belief. As was noted in Ref. [12], the micellar square-triangle order could cover a spherical surface too. Unfortunately, there haven't been any publications describing the synthesis of such a quasicrystalline system yet.

In our opinion, an experimental discovery of spherical QCs and their theoretical studies could be a breakthrough in modern physics, like the discoveries initiated by the works [6,17] as well as by the study of spherical colloidal crystals and colloidosomes [1,2,4]. The aim of this Letter is to consider the formation and peculiarities of spherical 2D dodecagonal QCs. We investigate regular spherical coverings, which are obtained from the parent planar quasicrystalline structures by a slight reconstruction of their order. We analyze how to minimize additional energy caused by a condition that the QC is enforced to assembly on a spherical template which could be a boundary between two liquid phases as is conventional for the self-assembly of spherical colloidal systems. This analysis allows predicting the more energetically favorable spherical QCs.

We construct the models of spherical structures from three dodecagonal tilings [18], which are shown in fig. 1 (a-c) and labeled as DTa, DTb and DTc, respectively. Note, that DTa was used to interpret structures of diblock copolymer [15]. The first experimentally observed QC micellar structure [12], nanoparticle superlattices [19], liquid polymer and colloidal quasicrystals [20] were interpreted in the frame of DTc. The structures of DTa and DTb are easily projected from six-dimensional space [21] and differ from each other by acceptance domains only [18]. DTc is usually constructed by the deflation method [22] since the tiling is characterized by the fractal acceptance domain [23] which is useless for application in the conventional projection method.

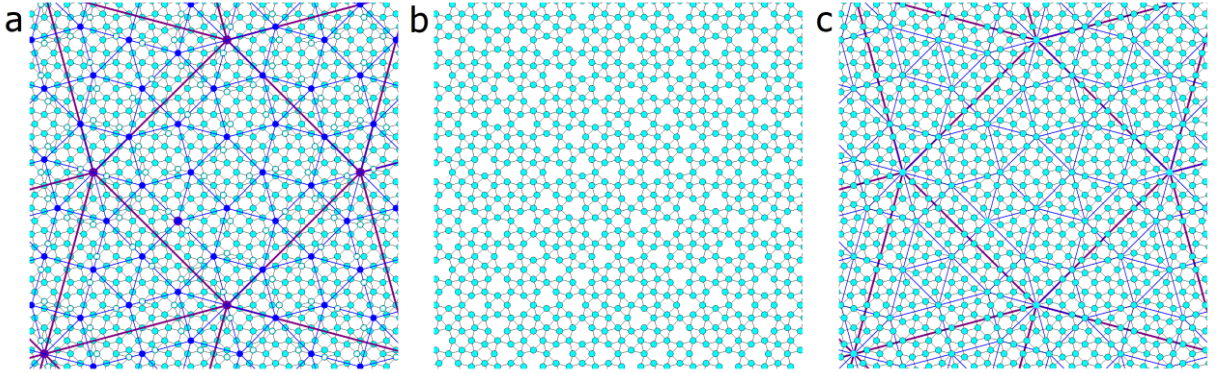

FIG. 1 (color online). Three widespread dodecagonal tilings. Since the tilings are constructed at the same nonzero value of Goldstone phason variable **v**, the most of their nodes coincide. (a) DTa is formed by squares, triangles and narrow rhombuses. (b) DTb contains distorted hexagons instead of rhombuses. (c) DTc consists of square and triangular tiles only. Identical inflated in $\tau$ and $\tau^2$ times square-triangle tilings depicted in panels (a) and (c) demonstrate the relation between DTa and DTc. The nodes of DTa shown by large (purple) circles form a minimal fragment of the square-triangle tiling inflated in $\tau^2$ times. Blue nodes form the tiling which appears after the first deflation.

Let us discuss some useful relations between the tilings demonstrated in Fig.1. Some nodes of DTa are shown in this figure by empty circles. They form the pairs separated by double tiling edges and are absent in DTb. In DTa the center of each double edge contains the node surrounded by a regular dodecagon. In DTc some of these nodes become cluster centers which we call 'wheels'. A wheel consists of 19 positions forming a regular hexagon inside the regular dodecagon. The wheels never overlap in DTc. However, an overlapping of imperfect wheels is possible in DTa. The distance between the centers of the overlapped wheels is equal to the diagonal of narrow rhombus and only one wheel from the pair appears in DTc after the order reconstruction. A choice of imperfect wheels to be included into DTc is difficult, because within their areas the chosen wheels should contain all rhombuses disappearing in the last tiling. We solve the choice problem and establish the relation between the DTa and DTc structures by the deflation method.

Note, that the distance between the nearest wheels in DTc is $\tau = \sqrt{3} + 2$ times longer than the length $a_{ed}$ of the tiling edge. DTa is self-similar with the same self-similarity coefficient. Therefore, in any finite fragment of DTa one can select a small number of nodes forming a minimal part of inflated square-triangular tiling with the length of edges equal to $a_{ed}\tau^n$. The areas with the size about $a_{ed}\tau^n$ around the selected nodes are characterized by a perfect $C_{12v}$ symmetry since the perpendicular coordinates of the selected nodes tend to zero as $\tau^{-n}$. Then, the consequent steps of deflation are performed and the wheels $\tau$ times smaller are inserted in the nodes of the tiling obtained at the previous step. At each step of deflation the internal hexagons inside the wheels can be rotated by π/6. To overcome this ambiguity we freeze each hexagon in the orientation which makes its vertices closer to the origin of the perpendicular space. Till the last step the deflated square-triangle tilings include only the nodes belonging to the original DTa. Finally, after deleting and switching a small number of positions the DTa is transformed into the perfect DTc.

As is it known some planar quasicrystalline tilings can be obtained by minimizing double-well interaction potentials [24]. Recent theoretical work [25] devoted to self-assembly of micellar QCs develops a similar approach. However, the tilings [25] do not correspond exactly to real micellar structures [12], where the micelles are not located in the tiling nodes but decorate the square and triangular tiles composing the packings. Besides, the experimental micellar structures are much more regular than the snapshots obtained [25]. So we do not use explicit energy minimization. Instead, a spherical quasicrystalline order is virtually obtained step by step. First, using geometrical and energetic reasons we search for the best polyhedron to transfer the planar QC structure onto a sphere. Second, we cover smoothly the polyhedron by the QC order. Conjugation of coverings at the adjacent faces induces a slight rearrangement of their structures. Finally, the polyhedron is buckled and forms the sphere.

Energy cost of this virtual mechanism includes three main contributions. They are: i) $E_{def}$ energy of topological defects arising in the polyhedron vertices. $E_{def}$ includes both the internal energy of defects and the energy of their interaction. ii) $E_{rec}$ energy of the phason strain, and iii) $E_{bck}$ energy related to the polyhedron buckling. Since we assume that QCs are enforced to assembly on a spherical template, the work of external forces (e.g. those of surface tension) should compensate the last term, which is neglected.

The full consideration of $E_{def}$ energy is difficult. To simplify the theory only the structures with the point topological defects are considered below. The exclusion of non-point (extended) defects greatly reduces the number of possible spherical QCs. One can clarify this fact using the example of spherical colloidal crystals, which are often modeled in the frame of Thomson problem [26]. The pioneer paper [27] devoted to the problem demonstrates that the number of metastable states with extremely close energies grows exponentially with the number of particles in the crystal. Topological defects with different internal structures are responsible for this phenomenon. Similar defects can also be typical for spherical QCs. Nevertheless, the extended defects in the spherical hexagonal order are located only in vicinity of spherical icosahedron vertices and the order is single connected. Therefore we hope that our consideration of the structures with the point disclinations only cannot impede the understanding of the basic features of the spherical QC order.

Now let us search for the best convex polyhedron minimizing $E_{def}$ energy. Note that the point disclinations located in the polyhadron vertices correspond to the sectors eliminated from the polyhedron net. For the net of any convex polyhedron the total angular value of the eliminated sectors is equal to $4\pi$. Since the smallest angular value $\Omega$ of the elininated sector for the dodecagonal case is $2\pi/12$, the total number of point disclinations is equal to 24. We assume that these defects repel each other to minimize $E_{def}$, as they do in spherical crystals [7]. Different repulsive pair potentials (including Coulomb one) lead to the defects located in the vertices of snub cube (SC), which is one of Archimedean polyhedrons. It corresponds to the fact that the SC is closer to the sphere than other polyhedrons with 24 vertices. Each of the SC vertices is sheared between four triangular and one square faces. The tilings shown in fig.1 include the tiles of the same shapes and the SC is the best polyhedron to be covered by dodecagonal tilings. However, a slight reconstruction of the quasicrystalline order is required. For a perfect matching of the adjacent faces to be achieved, the vertices of the SC net should coincide with the local 12-fold axes of the reconstructed tiling. After assembling the SC these axes turn into the local 11-fold ones of the spherical packing.

Now let us analyze $E_{rec}$ contribution which is nonzero for spherical QCs only. To transfer the crystalline hexagonal lattice onto a sphere the order matching at adjacent faces of the icosahedron is not required [7]. It is sufficient to construct a particular icosahedron net where the vertices coincide with the 6-fold axes of the hexagonal lattice. However, to match the local 12-fold axes of DTa and the vertices of the SC net without a tiling reconstruction is impossible. To determine the appropriate reconstruction we note that the triangular and square faces of the SC *can be smoothly covered by elementary cells of hexagonal and square periodic QC* *approximants*, respectively. A perfect matching of the adjacent cells is ensured by the coincidence of SC vertices with the local 12-fold symmetry axes of the approximants.

The approximants can appear from QCs due to the phase transitions inducing phason strains [28]. In our approach the phason strain appears due to the boundary conditions taking into account the order matching. To obtain $E_{rec}$ contribution we use these conditions and minimize the phason part

$$E_{rec} = \int \left[ K_1 (\partial_x v_x - \partial_y v_y)^2 + K_2 (\partial_x v_y + \partial_y v_x)^2 + K_3 \left( (\partial_x v_y - \partial_y v_x)^2 + (\partial_x v_x + \partial_y v_y)^2 \right) \right] dxdy \quad (1)$$

of the phonon-phason elastic free energy [29], where $K_i$ are the coefficients of phason elasticity, **v** stands for the phason strain field, and *{x,y}* are the components of radius-vector **R**.

Let the translation of hexagonal or square periodic approximants be a real space vector $\mathbf{Y}^{\|}$ being the parallel projection of some 6D translation **Y**. Then

$$n\mathbf{Y}^{\perp} - \mathbf{v}(n\mathbf{Y}^{\|}) = 0 , \quad (2)$$

where *n* is an arbitrary integer. Since energy (1) is quadratic, the field **v** satisfying condition (2) and minimizing this energy should be linear: $\mathbf{v}(\mathbf{R}) = \hat{m}\mathbf{R}$, where $\hat{m}$ is a phason strain tensor. Symmetry analysis shows that the tensor

$$m_{gex} = \begin{vmatrix} \gamma_1 & \gamma_2 \\ \gamma_2 & -\gamma_1 \end{vmatrix} \quad (3)$$

induces hexagonal approximants, while the tensor to obtain square approximants reads:

$$m_{sqr} = \begin{vmatrix} \alpha & \beta \\ -\beta & \alpha \end{vmatrix}. \quad (4)$$

To remove the symmetry-equivalent variants of approximants we direct vectors $\mathbf{Y}^{\|}$ from the vertex to the nodes inside the fundamental region, which is the $\pi/12$ sector located between two planes of $C_{12v}$ symmetry group (see Fig. 2).

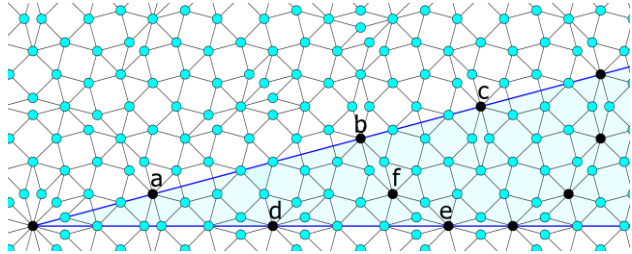

FIG. 2 (color online). Dodecogonal tiling DTa constructed at **v** = 0 condition. The 12-fold axis coincides with the vertex of the $\pi/12$ sector being the fundamental region of the tiling. To distinguish between more and less energetically favorable QCs (see the text) the nodes corresponding to the condition ($Y^{\perp} < 1/2$) are shown by black circles. The maximal value of $Y^{\perp}$ for nodes of the shown tiling is equal to 1.

Using the explicit form of vectors $\mathbf{Y}^{\|}$ and $\mathbf{Y}^{\perp}$ and taking into account Eq. (2) we calculate tensors (3-4): $\alpha = -(Y_x^{\perp} \cos\varphi + Y_y^{\perp} \sin\varphi)/Y^{\|}$; $\beta = (Y_y^{\perp} \cos\varphi - Y_x^{\perp} \sin\varphi)/Y^{\|}$; $\gamma_1 = (-Y_x^{\perp} \cos\varphi + Y_y^{\perp} \sin\varphi)/Y^{\|}$; $\gamma_2 = -(Y_y^{\perp} \cos\varphi + Y_x^{\perp} \sin\varphi)/Y^{\|}$, where $\cos\phi = Y_x^{\|}/Y^{\|}$. Then, Eq. (1) results in the phason strains energies $E_{tr}$ and $E_{sqr}$ of the triangular and square SC faces, respectively:

$$E_{trn} = \sqrt{3}K_1 (Y_x^{\perp} \cos\phi - Y_y^{\perp} \sin\phi)^2 + \sqrt{3}K_2 (Y_x^{\perp} \sin\phi + Y_y^{\perp} \cos\phi)^2 , \quad (5)$$

$$E_{sqr} = 4K_3 ((Y_x^{\perp})^2 + (Y_y^{\perp})^2) \quad (6)$$

The total reconstruction energy $32E_{trn}+6E_{sqr}$ of all SC faces is smaller for spherical QCs corresponding to smaller $Y^{\perp}$ values. The variables $Y^{\perp}$ and $\mathbf{Y}^{\|}$ are mutually dependent since they are the projections of the single vector **Y**. Therefore, QCs with the specific lengths $Y^{\|}$ of SC edges and the corresponding particular numbers of nodes are more energetically favorable (see Fig. 3). A similar phenomenon exists for spherical structures obtained in the frame of Tammes problem [30]. The packing density of spherical caps on a sphere increases for packings containing particular numbers of caps [31].

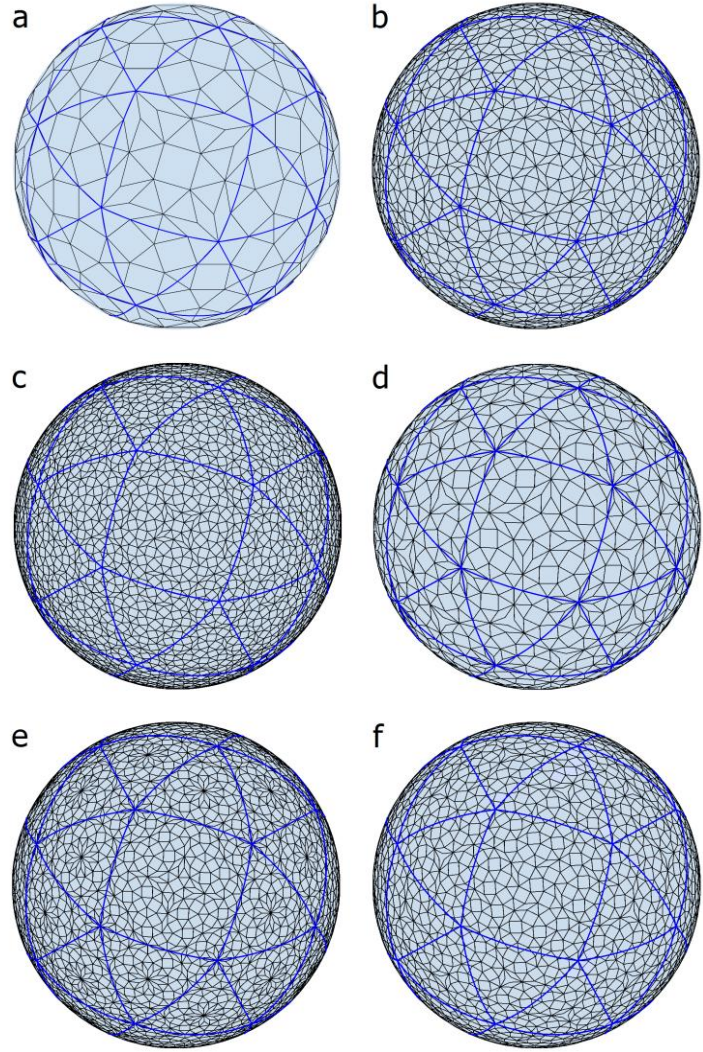

FIG. 3 (color online). Six energetically favorable spherical QCs obtained from the DTa tiling. Spherical SCs are drawn over the structures by blue lines. The QCs shown in panels (a-f) correspond to the $\mathbf{Y}^{\|}$ vectors connecting the vertex and the nodes (a-f) of the fundamental region (see Fig. 2). The numbers of nodes in the structures are 334, 2226, 3948, 1104, 3092, and 2562, respectively.

The relation between DTa and DTb (see Fig. 1) allows transforming the structures obtained into spherical tilings consisting of squares, triangles and oblate hexagons. However, construction of spherical QCs from DTc is more difficult. The 11-fold wheels forming the topological defects (see Fig. 3) cannot consist of squares and triangles only. The use of SC net for the order transfer induces additional defective tiles which increase the energetic cost of the process. In contrast, the use of icosahedron net allows obtaining the purely square-triangle order on the sphere. Of course, formation of 12 topological defects with double topological charge $2\Omega$ is less energetically favorable and the competition between the SC global organization and icosahedral one is possible. The competition result may be different in different systems. Below we consider only the icosahedral arrangement of the global spherical order, which is compatible with the purely square-triangle tilings. The construction of the related spherical QCs is obviously based on hexagonal periodic approximants of DTc.

Defect-free hexagonal and square approximants of DTc with the periodicity $T_n$ in $\tau^n$ times longer than $a_{ed}$ are known to be obtained by the deflation method. In order to apply it, the basic translations of approximants should be parallel to the tiling edges. In this case the vectors $\mathbf{Y}^\parallel$ with the length $T_n$ belong to the upper boundary of the fundamental region (see Fig. 2). If the basic translations of approximants are not parallel to the tiling edges then the defect-free packing of 12-fold wheels is impossible and the approximants include narrow rhombuses. However, since the hexagonal order admits overlapping of the wheels, there are perfect square-triangle approximants breaking the equality $T_n = a_{ed}\tau^n$. For example, in a series of hexagonal approximants the motif of the largest scale can be formed by three overlapping wheels with a common triangle. The periodicity $T_n'$ of these approximants reads $T_n' = a_{ed}(\tau-1)\tau^{n-1}$.

Moreover, any hexagonal approximant of DTa corresponding to the upper boundary of the fundamental region (see Fig. 2) can be reconstructed into the purely square-triangle structure with the same periodicity. Indeed, any vector $\mathbf{Y}^\parallel$ determining the approximant periodicity and lying along the upper boundary of the fundamental region can be expressed as the parallel projection (see Eqs. (1-2) in Ref. [21]) of the vector Y=<n,n,m,n-m,m-n,-m>, where *n* and *m* are arbitrary integers. This fact allows expressing the lengths of both projections $\mathbf{Y}^\parallel$ and $\mathbf{Y}^\perp$ in the following forms:

$$|Y^\parallel| = a_{ed}(\sqrt{3}p+q), \tag{7}$$

$$|Y^\perp| = \sqrt{3(4-\tau)}\,|\sqrt{3}p-q| \tag{8}$$

where *p = n-m* and *q = 2m-n*. The inverse relations are also integer: *n = 2p + q* and *m = p + q*. Eq. (7) is obviously compatible with the square-triangle order. The basic translation $\mathbf{Y}^\parallel$ consists of edges with different orientations. The *q* edges are parallel to $\mathbf{Y}^\parallel$ direction while *2p* edges are rotated by π/6 with respect to it. Like the DTa to DTc rearrangement, the reconstruction of DTa approximants into the DTc ones is based on the choice of imperfect wheels included into the square-triangle structure. Since the motives formed by these wheels in the DTa approximants are simple, the order reconstruction is not difficult.

Finally, all the above-obtained periodic approximants of DTc can be used to construct the perfect square-triangle spherical coverings. Some of them are shown in Fig. 4.

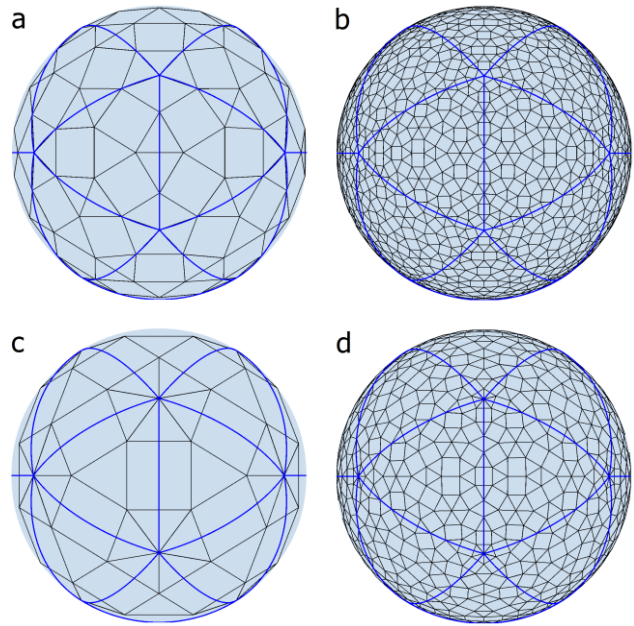

FIG. 4 (color online). Spherical structures related to the DTc tiling. Structures shown in panels (a-b) arise from $T_1$ and $T_2$ hexagonal approximants. Spherical tilings (c-d) are obtained from $T_1'$ and $T_2'$ approximants. The numbers of nodes in the shown spherical structures are 132, 1812, 72, 972, respectively. The term $E_{rec}$ makes the QCs demonstrated in panels (a-b) more energetically favorable than those shown in panels (c-d).

In conclusion, the synthesis of spherical QSs is very probable since the ordinary planar QCs due to the low-energy phason rearrangements and appearance of curvature-induced topological defects can smoothly cover spherical surfaces. In this Letter we consider the arrangement of spherical QCs originating from three widespread dodecagonal tilings. If the dodecagonal QC along with the square and triangular tiles admits a third structural element (narrow rhombus or flattened hexagon), then the sphere curvature induces 24 topological defects located near SC vertices and the resulting spherical packing is assembled on the basis of the SC net decorated by the same three types of tiles as the original planar tiling. The formation of the perfect square-triangular order on the sphere is also possible. However, in this case the topological defects should be located near the vertices of icosahedron. The physical properties and possible applications of dodecagonal spherical QCs may be even more interesting than those of different spherical crystals (from viral capsids to colloidosomes), which are actively studied now; we reserve these investigations for a future.

**Acknowledgements**

This work was supported by the RFBR grant 13-02-12085 ofi_m.

1. S. U. Pickering, J. Chem. Soc., Trans. **91**, 2001 (1907).
2. A.R. Bausch, M.J. Bowick, A. Cacciuto, A.D. Dinsmore, M.F. Hsu, D.R. Nelson, M.G. Nikolaides, A. Travesset, and D.A. Weitz, Science **299**, 1716 (2003).
3. T. Einert, P. Lipowsky, J. Schilling, M. Bowick and A.R. Bausch // Langmuir 21, 12076 (2005).
4. A.D. Dinsmore, Ming F. Hsu, M.G. Nikolaides, Manuel Marquez, A.R. Bausch, D.A. Weitz, Nature **298**, 1006 (2002).

5. R. Saito, G. Dresselhaus, and M. S. Dresselhaus, *Physical Properties of Carbon Nanotubes* (Imperial College Press, London, U.K., 1998)
6. S. Reich, C. Thomsen and J. Maultzsch, Carbon Nanotubes, Basic Concepts and Physical Properties (Wiley-VCH Verlag GmbH, Weinheim, Germany, 2004)
7. D.L. Caspar and A. Klug, Cold Spring Harbor Symp. Quant. Biol. **27**, 1 (1962).
8. J.E. Johnson and J.A. Speir, J. Mol. Biol. **269**, 665 (1997).
9. R. Twarock, J. Theor. Biol. **226**, 477 (2004).
10. O. V. Konevtsova, S. B. Rochal, and V. L. Lorman. Phys. Rev. Lett. **108**, 038102 (2012).
11. O. V. Konevtsova, S. B. Rochal, and V. L. Lorman, arXiv:1402.0201 (2014).
12. B. Zeng, G. Ungar, Y. S. Liu, V. Percec, A. E. Dulcey, J. K. Hobbs, Nature **428**, 157 (2004).
13. K. Hayashida, T. Dotera, A. Takano and Y. Matsushita, Phys. Rev. Lett. **98**, 195502 (2007).
14. J. Zhang, and F. S. Bates, J. Am. Chem. Soc. **134**, 7636 (2012).
15. S. Fischer, et al, Proc. Natl Acad. Sci. USA **108**, 1810 (2011).
16. C. Xiao, N. Fujita, K. Miyasaka, Y. Sakamoto, and O. Terasaki, Nature 4**87**, 349 (2012).
17. S. Iijima, Nature **354** (6348): 56 (1991).
18. F. Gahler, in: Ch. Janot, J.M. Dubois (Eds.), Quasicrystalline Materials, (World Scientific, Singapore, 1988).
19. D. V. Talapin, E. V. Schevchenko, M. I. Bodnarchuk, X. Ye, J. Chen, and C. B. Murray Nature **461**, 964 (2009).
20. T. Dotera, Isr. J. Chem.**51**, 1197 (2011).
21. Supplemental Material.
22. P. A. Stampfli, Helv. Phys. Acta. **59**, 1260 (1986).
23. M. Baake, R. Klitzing & M. Schlottmann, Physica A **191**, 554 (1992).
24. M. Engel, H.-R. Trebin, Phys. Rev. Lett. **98**, 225505 (2007).
25. T. Dotera, T. Oshiro, and P. Ziherl, Nature 506, 208 (2014).
26. J.J. Thomson, Phil. Mag. **7**, 237 (1904).
27. F. H. Stillinger & T. A. Weber, *Science*, **225**, 983-989 (1984).
28. V.E. Dmitrienko, M. Kleman, F. Mauri, Phys. Rev. B **60**, 9383 (1999).
29. J. E. S. Socolar, Phys. Rev. B 39, 10519 (1989).
30. P. M. L. Tammes, Recl. Trav. Bot. Neerl. **27**, 1 (1930).
31. N.J.A. Sloane, Sci. Am. **250**, 116 (1984).

**Supplemental Material**

**The construction of DTa and DTb dodecagonal tilings**

The positions of dodecagonal tilings are obtained by projecting integer nodes $\{n_i^j\}$, $i=0,1…5$ of 6D space $E$. This space contains three 2D irreducible subspaces transformed by $C_{12v}$ symmetry group. Subspaces $E^\parallel$ and $E^\perp$ are spanned by two different vector representations ($E_1$ and $E_5$, respectively) of this group. Projection of the space $E$ onto the first (parallel) subspace reads:

$$\mathbf{r}_j = \sum_{i=0}^{5} n_i^j \mathbf{a}_i, \qquad (1)$$

where $\mathbf{a}_i=(\cos(i\pi/6),\sin(i\pi/6))$ are the basis vectors and $i=0,1,2,3,4,5$. Expression for the projection upon the second (perpendicular) subspace is as follows:

$$\mathbf{r}_j^\perp = \sum_{i=0}^{5} n_i^j \mathbf{a}_i^\perp + \mathbf{v}, \qquad (2)$$

where $\mathbf{a}_i^\perp = (\cos(i5\pi/6), \sin(i5\pi/6))$ are the basis vectors of the perpendicular space $E^\perp$, $i=0,1,2,3,4,5$; and $\mathbf{v}$ is a phason Goldstone degree of freedom. The node (j) is included into the tiling if the vector $\mathbf{r}_j^\perp$ belongs to the acceptance domain, which is the conjugation (for DTa) or intersection (for DTa) of the two superimposed regular hexagons [20]. These hexagons are relatively rotated by $\pi/6$, and their vertexes coincide with the twelve vectors $\pm \mathbf{a}_i^\perp$. The tiling edges are the parallel projections of 6D translations which are symmetry equivalent to the initial one <1,1,1,1,0,0>. The length $a_{ed}$ of the tiling edges is equal to $\sqrt{2}+\sqrt{\tau}$. Note that the acceptance domain of DTc has a fractal form [23]. This domain is useless to project this tiling from a high dimensional space and DTc can be obtained from DTa as it is described in the Letter.

The third subspace is spanned by the $E_4$ irreducible representation. Its basis has the following form:

$$\chi_1 = n_0^j - n_2^j + n_4^j, \quad \chi_2 = n_1^j - n_3^j + n_5^j \qquad (3)$$

Different possible values of the integer variables $\chi_1$ and $\chi_2$ lead to quite similar structures. We take here $\chi_1=0$ and $\chi_2=0$ since in this case if $v = 0$ than the origin of space $E$ is projected onto the 12-fold global axis of the tiling (see Fig. 2).